\documentstyle[sprocl,epsfig,rotate,amssymb]{article}

\def\be{\begin{equation}}
\def\ee{\end{equation}}
\def\bea{\begin{eqnarray}}
\def\eea{\end{eqnarray}}
\begin{document}

\title{NON-PERTURBATIVE CHIRAL CORRECTIONS FOR LATTICE QCD} 

\author{A. W. THOMAS, D. B. LEINWEBER }

\address{Department of Physics and Mathematical Physics and \\
Special Research Centre for the Subatomic Structure of Matter\\
University of Adelaide\\ Adelaide SA 5005, AUSTRALIA \\
E-mail: athomas,dleinweb@physics.adelaide.edu.au}

\author{D. H. LU}

\address{Department of Physics, National Taiwan University,\\
Taipei, TAIWAN\\
E-mail: dhlu@phys5.phys.ntu.edu.tw}


\maketitle
\vspace{-9.9cm}
\begin{flushright}
{\footnotesize Invited talk presented at NEWS 99} \hspace{1cm} \\
{\footnotesize University of Osaka, March 9-12, 1999} \\
{\footnotesize ADP-99-18/T360 \hspace{2cm}}
\end{flushright}
\vspace{8.9cm}
\abstracts{We explore the chiral aspects of extrapolation of
observables calculated within lattice QCD, using the nucleon magnetic
moments as an example. Our analysis shows that the biggest effects of
chiral dynamics occur for quark masses corresponding to a pion mass
below 600 MeV. In this limited range chiral perturbation theory is not
rapidly convergent, but we can develop some understanding of the
behaviour through chiral quark models. This model dependent analysis
leads us to a simple Pad\'e approximant which builds in both the limits
$m_\pi \rightarrow 0$ and $m_\pi \rightarrow \infty$ correctly and
permits a consistent, model independent extrapolation 
to the physical pion mass which
should be extremely reliable.}

\section{Introduction}

At present the only known way to
calculate the properties of QCD directly is through the formulation of
lattice gauge theory on a discrete space-time lattice.
While the lattice formulation of QCD is well
established \cite{lattice98}, there have recently been a number of
exciting advances in lattice action improvement
which are greatly facilitating the reduction of systematic uncertainties
associated with the finite lattice volume and the finite lattice
spacing.  However, direct simulation of QCD for light current quark
masses, near the chiral limit, remains computationally intensive. 
In particular, present lattice calculations of masses in
full QCD are limited to equivalent pion masses of order 500 MeV or more
-- although there have been recent preliminary results from CP-PACS
using improved actions at a mass as low as 300 MeV \cite{2}. For nucleon
magnetic moments the situation is much worse, with the best calculations
still being made within quenched QCD for pion masses greater than 600
MeV. This is clearly beyond the range of validity of chiral perturbation
theory ($\chi$PT).

\newpage

In view of this situation, the present approach of calculating 
the properties of QCD using
quark masses away from the chiral regime and extrapolating to the
physical world is likely to persist for the foreseeable future.
It is therefore vital to understand the quark mass dependence of 
hadronic observables calculated on the lattice and how to connect these
calculations with the physical world. A major difficulty in this
endeavour is the rapid rise of the pseudoscalar mass for small increases
in the quark mass away from the chiral limit -- with $m_\pi^2 \propto
\bar{m}$, where $\bar{m}$ is the light quark mass.

Historically, lattice results were often linearly extrapolated with
respect to $m_\pi^2$, particularly in exploratory
calculations.  More recently the focus has turned to chiral
perturbation theory ($\chi$PT), which provides predictions for the
leading nonanalytic quark-mass dependence of observables in terms of
phenomenological parameters \cite{golterman94,labrenz94}.
Indeed, it is now relatively standard to use $\chi$PT in the
extrapolation of lattice simulation data for hadron masses and decay
constants \cite{lattice98}. On the other hand, earlier attempts
to apply $\chi$PT predictions for the quark-mass
dependence of baryon magnetic
moments failed, as the higher order terms of the chiral expansion
quickly dominate the truncated expansion as one moves away from the
chiral limit.  To one meson loop, $\chi$PT expresses the nucleon
magnetic moments as \cite{jenkins93}
\begin{equation}
\mu_N = \mu_0 + c_1 \, m_\pi + c_2 \, m_\pi^2 \log m_\pi +
c_3 \, m_\pi^2 + \cdots \, ,
\end{equation}
where $\mu_0$ and $c_3$ are fitted phenomenologically and $c_1$ and
$c_2$ are predicted by $\chi$PT.  The $m_\pi^2 \log m_\pi$ term
quickly dominates as $m_\pi$ moves away from the chiral limit, making
contact with the lattice results impossible.

As a result of these early difficulties, lattice QCD results for baryon
magnetic moments\cite{dblOctet,wilcox92,dong97} remain predominantly
based on linear quark mass (or $m_\pi^2$) extrapolations of the moments
expressed in natural magnetons.  This approach systematically
underestimates the
measured moments by 10 to 20\%.  Finite lattice volume and spacing
errors are expected to be some source of systematic error.  However,
$\chi$PT clearly indicates the linear extrapolation of the simulation
results is also suspect.  It is therefore imperative to find a method
which can bridge the void between the realm of $\chi$PT and lattice
simulations.

We report such a method, which provides predictions for
the quark mass (or $m_\pi^2$) dependence of nucleon magnetic moments
well beyond the chiral limit -- for more details we refer to
Ref.~\cite{Us}. The method is motivated by studies based
on the cloudy bag model (CBM), a chiral quark model which preserves the
correct leading non-analytic behaviour of chiral perturbation theory
while providing what should be a fairly reliable transition to the
regime of large pion mass. These studies suggest a Pad\'e approximant
which incorporates both the leading nonanalytic structure of $\chi$PT
and Dirac-moment mass dependence in the heavy quark-mass regime. We
apply this method to the existing lattice data as an illustration of how
important such an approach will be to the analysis of future lattice QCD
calculations.

\section{The Cloudy Bag Model}

The linearized CBM Lagrangian, with pseudoscalar pion-quark coupling
(to order $1/f_\pi$), is given by \cite{CBM,thomas84}
\begin{eqnarray}
\protect{\cal L}
&=& \left[ \overline q (i\gamma^\mu \partial_\mu-m_q)q -
B\right]\theta_V
- {1\over 2}\overline q q \delta_S \nonumber \\
&& + {1\over 2} (\partial_\mu \pi)^2
- {1\over 2} m^2_\pi \pi^2
- {i\over 2f_\pi} \overline q \gamma_5 \tau \cdot
\pi q \delta_S,
\label{LAG}
\end{eqnarray}
where $B$ is the bag constant, $f_\pi$ is the $\pi$ decay constant,
$\theta_V$ is a step function (unity inside the bag volume and
vanishing outside) and $\delta_S$ is a surface delta function.  In a
lowest order perturbative treatment of the pion field, the quark wave
function is not affected by the pion field and is simply given by the
MIT bag model.  Our calculation is carried out in
the Breit frame with the center-of-mass correction for the bag
performed via Peierls-Thouless projection.  The detailed formulas for
calculating nucleon electromagnetic form factors in the CBM are given
in Ref.~\cite{lu98}.

In the CBM, a baryon is viewed as a superposition of a bare quark
core and bag plus meson states.  Both the quark core and the meson
cloud contribute to the baryon magnetic moments. These two sources
are balanced around a bag radius, $R = 0.7 - 1.1$ fm \cite{MM}.
A large bag radius suppresses the contributions from the pion cloud, and
enhances the contribution from the quark core.
The CBM reproduces the leading non-analytic behavior of $\chi$PT, which
has as its origin the process shown in Fig. 1(c), along with the other
contributions in the model.
\begin{figure}[ht]
%\vspace{1.5cm}
\centering{\
\epsfig{file=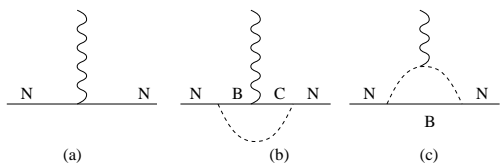,height=2cm}
%\vspace*{1.0cm}
\caption{Schematic illustration of the processes included in the CBM
calculation.  $B$ and $C$ denote intermediate state baryons and
include $N$ and $\Delta$.}
\label{fig1.ps}}
\end{figure}

{}For the $\pi NN$ vertex, 
%instead of the generic form 
we take a
phenomenological, monopole form, $u(k)=(\Lambda^2 - \mu^2)/(\Lambda^2 +
k^2)$, where $k$ is the loop momentum and $\Lambda$ is a cut-off
parameter. As current lattice simulations indicate that $m_\pi^2$ is
approximately proportional to $m_q$ over a wide
range of quark masses~\cite{cppacs97}, we scale the mass of the quark
confined in the bag as $m_q= \left(m_\pi/\mu\right)^2 m_q^{(0)}$, with
$m_q^{(0)}$ being the current quark mass corresponding to the physical
pion mass ($\mu$). In order to obtain a first idea of the behaviour within the
model between the lowest mass lattice point and the experimental data,
the parameters $R_0$, $\Lambda$ and $m_q^{(0)}$ were tuned to reproduce
the experimental moment while accommodating the lattice data.
It turns out that the best fits are obtained with $m_q^{(0)}$
in the range 6 to 7 MeV for a bag radius of 0.8fm and $\Lambda$ of order
600-700 MeV, which are all quite satisfactory. 

\begin{figure}[ht]
\centering{\
\epsfig{file=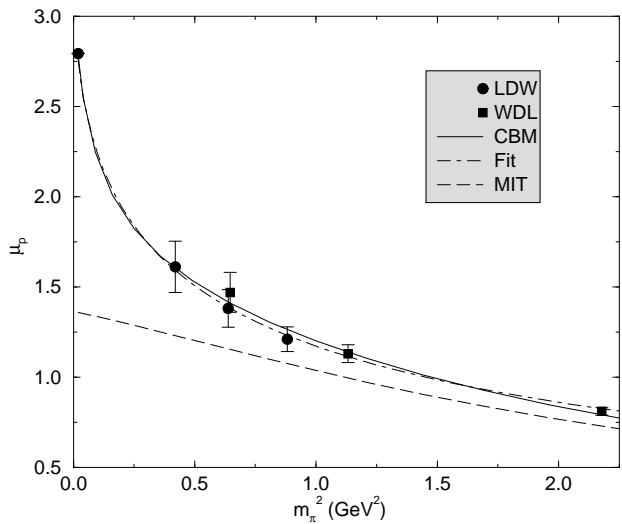,height=7cm}
\caption{The proton magnetic moment as calculated in lattice QCD
($\bullet$ LDW Ref.\ \protect\cite{dblOctet},
\protect $\scriptstyle\blacksquare$ WDL Ref.\ \protect\cite{wilcox92}), the
cloudy bag model (CBM) and the MIT bag model (MIT). 
Also illustrated is a fit of the simple analytic form given in 
Eq.~(\protect\ref{fit}) to the CBM results.  The point at the physical
value of $m_\pi^2$ is the experimental measurement and is used to
constrain the parameters of the CBM.}
\label{proton.ps}}
\end{figure}
Inspection of the results of this calculation, in Fig. \ref{proton.ps},
show clearly that the pion cloud contribution 
to the nucleon magnetic moments decreases very
quickly, becoming quite small for large quark masses -- especially in
the range corresponding to the current lattice calculations. In
particular, the total pionic correction at the first lattice data point
(for $m_\pi$ around 600 MeV) is of the order of only 10-15\% of the 
total. Since the present lattice data is based on a quenched
calculation, which gives incorrect chiral contributions, this is quite
good news. It suggests that in using this data as the basis for a
correct chiral extrapolation to the physical mass quenching should not
induce a major error.

It is also clear from Fig. \ref{proton.ps} that below $m_\pi$ = 600 MeV
the behaviour of the magnetic moment is highly non-linear. One clearly
needs to account for such behaviour in a reliable manner if we are to 
make believable extrapolations of the lattice data. Rather than rely on a
model dependent extrapolation method, we chose to investigate whether it
was possible to find a simple phenomenological form, {\em with a sound
physical basis}, which could do the job. The successful conclusion to
that search is described next.

\section{Encapsulating Formula}

After considerable effort we found that the following simple Pad\'e
approximant, which builds in both the linear behaviour in $m_\pi$ as
$m_\pi \rightarrow 0$ (i.e., a square root branch point in $\bar{m}$)
and the Dirac moment for large $m_\pi^2$ (i.e., $\mu_N \sim 1/\bar{m}$),
was able to reproduce the behaviour found in the CBM calculations:
\begin{equation}
\mu_N(m_\pi) = {\mu_N^{(0)} \over 1 + \frac{\alpha}{\mu_N^{(0)}}
m_\pi + \beta m_\pi^2}
\, .
\label{fit}
\end{equation}
The fit for the proton case is also shown in Fig. 2. Even more
remarkable is that the fit parameter $\alpha$ turned out to be quite
close to the value required by $\chi$PT (e.g., $\alpha = 4.54$ for the
proton, compared with $\alpha = 4.41$ in $\chi$PT).

\begin{figure}[ht]
\centering{\
\rotate{\epsfig{file=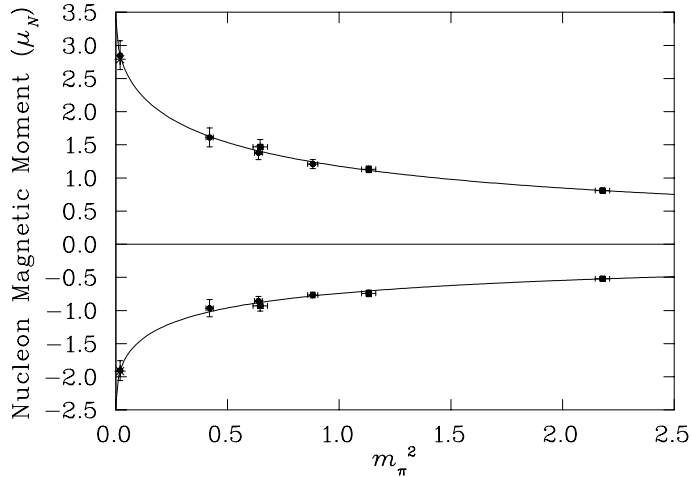,height=9cm}}
\caption{Extrapolation of lattice QCD magnetic moments 
for the proton (upper) and
neutron (lower) to the chiral limit.  The curves illustrate a
two parameter fit of Eq.~(\protect\ref{fit}) to the simulation data in which
the one-loop corrected chiral coefficient of $m_\pi$ is taken from
$\chi$PT.  The experimentally measured moments are indicated by
asterisks.}
\label{NucleonMomFit}}
\end{figure}
This result encourages us to propose that Eq.(\ref{fit}), with $\alpha$
taken from $\chi$PT, should be used as the method of extrapolation in
future analyses of lattice data for nucleon magnetic moments -- and
suitably generalized, for all members of the baryon octet. As an
example, we show in Fig.3 the result of two parameter fits
to the proton and neutron data. 
The nucleon magnetic moments at the physical pion mass, obtained from
this extrapolation, are
\begin{equation}
\mu_p = 2.85(22)\ \mu_N \quad \mbox{and} \quad \mu_n = -1.90(15)\ \mu_N
,
\end{equation}
which agree surprisingly well with the experimental measurements, 2.793
and $-$1.913 $\mu_N$ respectively.  We note that the data required
to do a fit of the lattice results in which covariances are taken into
account is no longer available.  As such, the uncertainties quoted
here should be regarded as indicative only.

\section{Conclusion}

We have explored the quark mass dependence of nucleon
magnetic moments.  Quark masses beyond the region appropriate to chiral
perturbation theory have been explored using the cloudy bag model (CBM) which
reproduces the leading nonanalytic behaviour of $\chi$PT while modeling 
the internal structure of the hadron under investigation.  We find that
the predictions of the CBM are succinctly described by a simple
formula which reproduces the leading nonanalytic behavior of $\chi$PT
in the limit $m_\pi \to 0$ {\em and} provides the anticipated Dirac moment
behavior in the limit $m_\pi \to \infty$. As an example we applied this
encapsulating formula to the existing lattice data for nucleon magnetic
moments, leading to surprisingly accurate predictions in comparison with
the observed values. It  will be interesting to see how the fit
parameters change as finite volume and lattice spacing artifacts are
eliminated in future simulations and whether the level of agreement seen
in this investigation is maintained or improved. We strongly advocate
the use of the Pad\'e approximant 
given in Eq.~(\ref{fit}) in future lattice QCD
investigations of octet baryon magnetic moments.

\section*{Acknowledgements}
This work was supported by the Australian Research Council.

\end{document}